\documentclass[10pt,twocolumn,letterpaper]{article}

\usepackage{wacv}
\usepackage{times}
\usepackage{epsfig}
\usepackage{graphicx}
\usepackage{amsmath}
\usepackage{amssymb}
\usepackage{mathtools}
\DeclarePairedDelimiter{\norm}{\lVert}{\rVert}

\wacvfinalcopy 

\def\wacvPaperID{677} 

\ifwacvfinal\fi
\begin{document}

\title{Efficient Video Super-Resolution through Recurrent Latent Space Propagation}

\author{Dario Fuoli \hspace{2cm} Shuhang Gu \hspace{2cm} Radu Timofte \\
Computer Vision Lab, ETH Zurich, Switzerland\\
{\tt\small \{dario.fuoli, shuhang.gu, radu.timofte\}@vision.ee.ethz.ch}
}

\maketitle
\ifwacvfinal\thispagestyle{empty}\fi

\begin{abstract}

With the recent trend for ultra high definition displays, the demand for high quality and efficient video super-resolution (VSR) has become more important than ever.
Previous methods adopt complex motion compensation strategies to exploit temporal information when estimating the missing high frequency details.
However, as the motion estimation problem is a highly challenging problem, inaccurate motion compensation may affect the performance of VSR algorithms.
Furthermore, the complex motion compensation module may also introduce a heavy computational burden, which limits the application of these methods in real systems. 
In this paper, we propose an efficient recurrent latent space propagation (RLSP) algorithm for fast VSR.
RLSP introduces high-dimensional latent states to propagate temporal information between frames in an implicit manner.
Our experimental results show that RLSP is a highly efficient and effective method to deal with the VSR problem.
%
We outperform current state-of-the-art method \cite{duf} with over 70$\times$ speed-up.
%

%

\end{abstract}

\section{Introduction}
\begin{figure}
\begin{center}
\includegraphics[width=\linewidth]{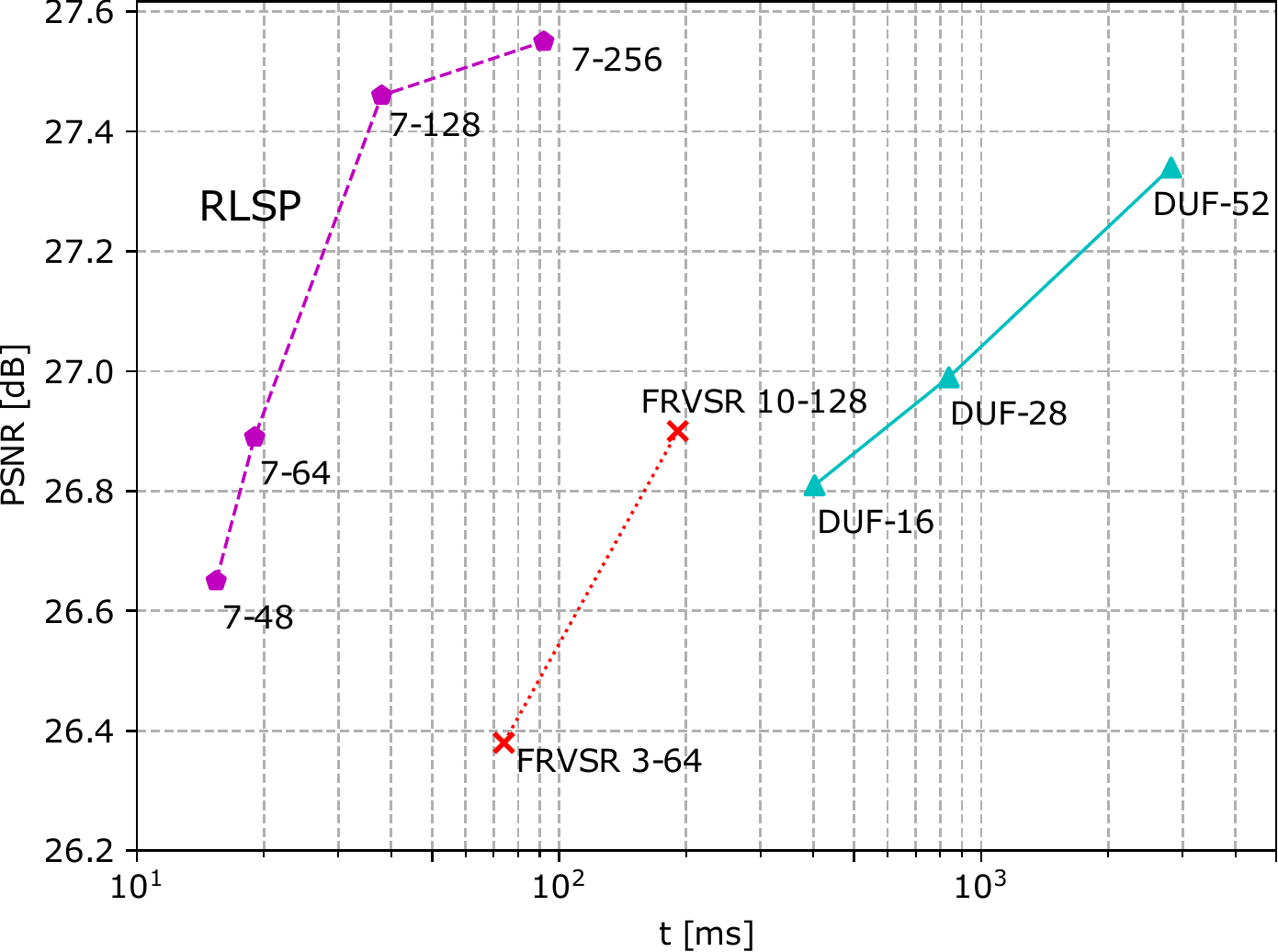}
\end{center}
   \caption{Quantitative comparison of PSNR values on Vid4 and computation times to produce a single Full HD (1920x1080) frame with other state-of-the-art methods FRVSR~\cite{frvsr} and DUF~\cite{duf}.}
\label{fig:t_vs_acc}
\end{figure}

Super-resolution aims to obtain high-resolution (HR) images from its low-resolution (LR) observations.
It provides a practical solution to enhance existing images as well as alleviating the pressure of data transportation.
One category of methods~\cite{srcnn, vdsr, srgan, edsr, pyramid, back_proj} takes a single LR image as input. The single image super-resolution (SISR) problem has been intensively studied for many years and is still an active topic in the area of computer vision.
Another category of approaches~\cite{hyun2018spatio, frvsr, duf, uhd_sparse, borsoi_robust, multi_mem}, \ie video super-resolution (VSR), takes LR video as input.
In contrast to SISR methods, which can only rely on natural image priors for estimation of high resolution details,
VSR exploits temporal information for improved recovery of image details.

A key issue to the success of VSR algorithms, is how to take full advantage from temporal information~\cite{vsroverview}.
In the early years, different methods have been suggested to model the subpixel-level motion between LR observations, including the Bilateral prior~\cite{farsiu} and Bayesian estimation model~\cite{bayes} have been adopted to solve the VSR problem.
In recent years, the success of deep learning in other vision tasks inspired the research to apply convolutional neural networks (CNN) also to VSR.
Following a similar strategy, already adopted in conventional VSR algorithms, most of existing deep learning based VSR methods divide the task into two sub-problems: motion estimation and the following compensation procedure.
In the last several years, a large number of elaborately designed models have been proposed to capture the subpixel motion between input LR frames.
However, as subpixel-level alignment of images is a highly challenging problem, these types of approaches may generate blurred estimations, when the motion compensation module fails to generate accurate motion estimation.
Furthermore, the complex motion estimation and compensation modules are often computationally expensive, which makes these methods unable to handle HR video in real time.

To address the accuracy issue, Jo~\etal~\cite{duf} proposed dynamic upsampling filters (DUF) to perform VSR without explicit motion compensation.
 In their solution, motion information is implicitly captured with dynamic upsampling filters and the HR frame is directly constructed by local filtering of the center input frame.
 Such an implicit formulation avoids conducting motion compensation in the image space and helps DUF to obtain state-of-the-art VSR results.
 However, as DUF needs to estimate dynamic filters in each location, the algorithm suffers from heavy computation as well as putting a burden on memory for processing large size images.

 In this paper, we propose a recurrent latent space propagation (RLSP) method for efficient VSR.
 RLSP follows a similar strategy as FRVSR~\cite{frvsr}, which utilizes a recurrent architecture to avoid processing LR input frames multiple times. 
 In contrast to FRVSR, which adopts explicit motion estimation and warping operations to exploit temporal information, RLSP introduces high dimensional latent states to propagate temporal information in an implicit way.
 %

In Fig.~\ref{fig:t_vs_acc}, we present the trade-off between runtime and accuracy (average PSNR) for state-of-the-art VSR approaches on the Vid4 dataset~\cite{bayes}.
The proposed RLSP approach achieves a better balance between speed and performance than the competing methods.
RLSP achieves about 10$\times$ and 70$\times$ speed-up over the methods FRVSR and DUF, respectively, while maintaining similar accuracy.
Furthermore, despite its efficiency, by utilizing more filters in our model, RLSP can also be pushed to pursue state-of-the-art VSR accuracy.
Our model RLSP 7-256 achieves the highest PSNR on the Vid4 benchmark.

\begin{figure*}
\begin{center}
\includegraphics[width=1\linewidth]{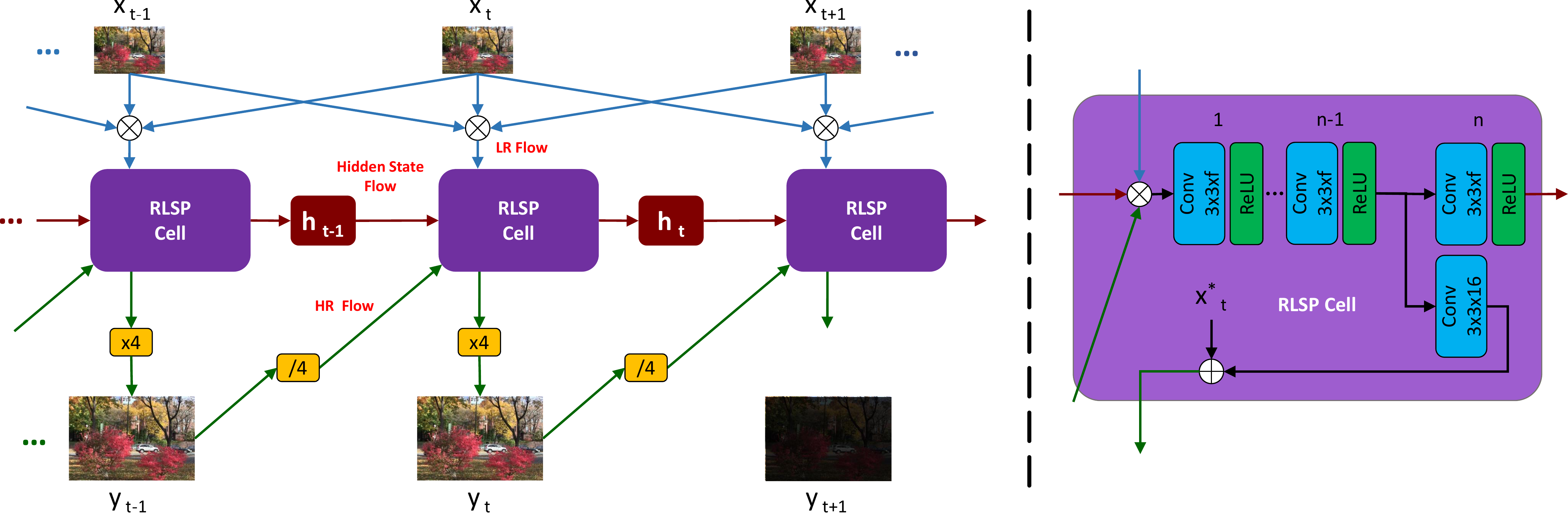}
\end{center}
   \caption{RLSP Architecture. The recurrent cell is shown at time $t$, $\otimes$ denotes concatenation along the channel dimension, $\oplus$ denotes element-wise addition. Information is propagated over time through hidden state $h$ and feedback.}
\label{fig:model}
\end{figure*}


\section{Related Work}
\noindent
\textbf{Single Image Super-Resolution}
\newline
With the rise of deep learning, especially convolutional neural networks (CNN)~\cite{imagenet}, learning based super-resolution models have shown to be superior in terms of accuracy compared to classical interpolation methods, such as bilinear and bicubic interpolation and similar approaches. One of the earliest methods to apply convolution for super-resolution is SRCNN, proposed by ~\cite{srcnn}. SRCNN uses a shallow network of only 3 convolutional layers. VDSR~\cite{vdsr} shows substantial improvements by using a much deeper network of 20 layers combined with residual learning. In order to get visually more pleasing images, photorealistic and natural looking, the accuracy to the ground truth is traded off by method such as SRGAN~\cite{srgan}, EnhanceNet~\cite{enhancenet}, and~\cite{photogan, dahl, lucas} that introduce alternative loss functions~\cite{gan} to super-resolution. An overview of recent methods in the field of SISR is provided by~\cite{timofte_challenge}.

\noindent
\textbf{Video Super-Resolution}
\newline
Super-resolution can be generalized from images to videos. Videos additionally provide temporal information among frames, which can be exploited to improve interpolation quality. Non-deep learning video super-resolution problems are often solved by formulating demanding optimization problems, leading to slow evaluation times~\cite{belekos, farsiu, bayes}.

Many deep learning based VSR methods are composed of multiple independent processing pipelines, motivated by prior knowledge and inspired by traditional computer vision tools. To leverage temporal information, a natural extension to SISR is combining multiple low-resolution frames to produce a single high-resolution estimate~\cite{liao, endtoend, sparse_rep}. 
Kappeler~\etal~\cite{kappeler} combine several adjacent frames. Non-center frames are motion compensated by calculating optical flow and warping towards the center frame. All frames are then concatenated and followed by 3 convolution layers. 
Tao~\etal~\cite{tao} produce a single high-resolution frame $y_t$ from up to 7 low-resolution input frames $x_{t-3:t+3}$. First, motion is estimated in low resolution and a preliminary high-resolution frame is computed through a subpixel motion compensation layer. The final output is computed by applying an encoder-decoder style network with an intermediate convolutional LSTM~\cite{lstm} layer. 
Liu~\etal~\cite{liu} calculate multiple high-resolution estimates in parallel branches, each processing an increasing number of low-resolution frames. Additionally, a temporal modulation branch computes weights according to which the respective high-resolution estimates are aggregated, forming the final high-resolution output.
Caballero~\etal~\cite{caballero} extract motion flow maps between adjacent frames and center frame $x_t$. The frames are warped according to the flow maps towards frame $x_t$. These frames are then processed with a spatio-temporal network, by either direct concatenation and convolution, gradually applying several convolution and concatenation steps or applying 3D convolutions~\cite{3Dconv}.
Jo~\etal~\cite{duf} propose DUF, a network without explicit motion estimation. Dynamic upsampling filters and residuals are calculated from a batch of adjacent input frames. The center frame is filtered and added with the residuals to get the final output.

A more powerful approach to process sequential data like video, is to use recurrent connections between time steps. Methods using a fixed number of input frames are inherently limited by the information content in those frames. Recurrent models however, are able to leverage information from a potentially unlimited number of frames.
Sajjadi~\etal~\cite{frvsr} use an optical flow network, followed by a super-resolution network. Optical flow is calculated between $x_{t-1}$ and $x_t$ to warp the previous output $y_{t-1}$ towards $t$. The final output $y_t$ is calculated from the warped previous output and the current low-resolution input frame $x_t$. The two networks are trained jointly.
Huang~\etal~\cite{brcn} propose a bidirectional recurrent network using 2D and 3D convolutions with recurrent connections between time steps. A forward pass and a backward pass are combined to produce the final output frames. Because of its nature, this method can not be applied online.

RLSP does not rely on a dedicated motion compensation module and instead introduces a recurrent hidden state, to efficiently leverage temporal information implicitly.

\section{Method}
Video super-resolution (VSR) maps a LR video $x$ to a HR video $y$ by a given scaling factor $r$. At time $t$, a single frame $y_t \in \mathbb{R}^{rH\times rW \times C}$ represents the reconstructed HR frame of $x_t \in \mathbb{R}^{H\times W \times C}$, where $H$ and $W$ are spatial dimensions and $C$ is the number of color channels.
In contrast to SISR, the temporal dimension provides additional information, which can be leveraged when generating a single frame $y_t$. As a natural choice for sequential data, we therefore propose a recurrent neural network (RNN). The model is fully defined by its cell, illustrated in Fig.~\ref{fig:model}. 

As done in many non-recurrent methods, we feed several adjacent LR frames $x_{t-1:t+1}$ to produce $y_t$. These frames are concatenated along the channel axis, together with the recurrent inputs $h_{t-1}$ and $y_{t-1}$. To concatenate and align the previous HR output $y_{t-1}$ with the LR tensors, $y_{t-1}$ is shuffled down by the scaling factor $r$, see Sec.~\ref{sec:shuffling}. The combined input is then fed to $n$ convolution layers with ReLU activation function. In the last stage, the hidden state $h_t$ for the next iteration and the HR output's residuals in LR space are produced. The residuals are added with the nearest neighbor interpolated frame $x^*_t$ represented in LR space and shuffled up by scaling factor $r$, to finally generate the output $y_t$ (see Sec.~\ref{sec:residual_learning}). All input frames are in RGB color space, while the output represents the brightness channel Y of YCbCr color space. Chroma channels are upscaled separately with bicubic interpolation. All our models are trained with a scaling factor of $r=4$.

The model is a recurrent, fully convolutional network. It is therefore not limited to a fixed input size and can accommodate video data of any dimensions. With the exception of higher level contextual information (along the spatial and temporal axes), super-resolution is a highly locality based interpolation problem. Thus, a convolutional neural network is a sensible choice. Since the receptive field grows with the number of convolution layers, the network is still able to detect complex structure across an extended neighborhood. In order to optimize information flow, much care is taken to keep local alignment throughout the network and is achieved by using operations which keep local integrity, also see Sec.~\ref{sec:shuffling} and Sec.~\ref{sec:residual_learning}. 
We aim for efficiency and therefore use a fixed number of $n=7$ layers, the filter's spatial dimensions are set to $3 \times 3$ for all models. The number of filters $f$ is adapted per model.

In the following sections, the model's core elements are discussed in more detail.

\subsection{Shuffling}
\label{sec:shuffling}

To realize the mapping from LR to HR, the spatial dimensions need to be transformed at some point in the model. The most relevant processing is kept in LR and spatial expansion is executed at the last stage of the processing chain. Because the output is fed back, an inverse transformation from HR to LR is applied. Shuffling performs these transformations by reducing the channel dimension $Z$ of a tensor $t$ with factor $r^2$ and extending both spatial dimensions $H$ and $W$ with factor $r$ and vice versa for the inverse transformation. Since shuffling is a bijective transformation these operations are reversible:
\begin{equation}
t^{LR} \in\mathbb{R}^{H\times W\times Z} \quad \xrightarrow{\times r} \quad t^{HR} \in\mathbb{R}^{rH\times rW\times Z/{r^2}}
\end{equation}
\begin{equation}
t^{HR} \in\mathbb{R}^{H\times W\times Z} \quad \xrightarrow{/r} \quad t^{LR} \in\mathbb{R}^{H/r\times W/r\times r^2Z}
\end{equation}
To get a single channel HR output image with upscaling factor $r=4$, the LR tensor's channel dimension needs to be $Z=16$. Therefore, the last layer has 16 filters. A very important characteristic of this transformation is that it retains local integrity. All pixels along the channel dimension in LR are rearranged in their corresponding local HR interpolation area. This enables a smooth localized information flow from LR input to HR output.
The shuffling operation has been adopted in previous works \cite{ESPCN, frvsr} to change the spatial resolution of image/feature maps. 
\begin{figure}
\begin{center}
\includegraphics[width=0.8\linewidth]{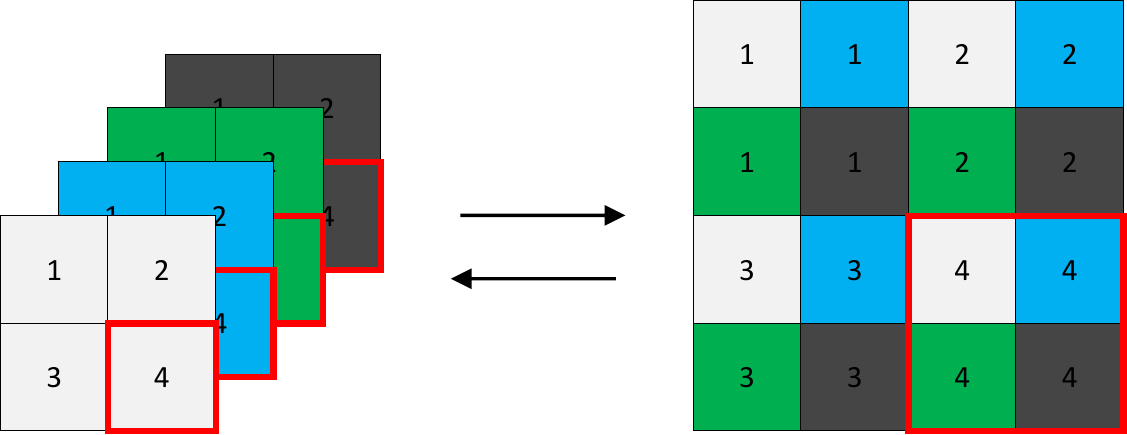}
\end{center}
   \caption{Shuffling: $t^{LR} \in\mathbb{R}^{2\times 2\times 4} \xrightarrow{\times 2} t^{HR} \in\mathbb{R}^{4\times 4\times 1}$.}
\label{fig:latentspace}
\vspace{-0.14cm}
\end{figure}

\subsection{Residual Learning}
\label{sec:residual_learning}
Reducing the sample rate inherently leads to loss of frequency components above the Nyquist frequency. Due to the lower Nyquist frequency in LR space, the main information loss occurs in the spatial high-frequency components. Low-frequency components below the Nyquist frequency can be fully retained when downsampling. This a priori knowledge is used by introducing a residual connection directly from the LR input frame $x_t$ to the output frame $y_t$. First, $x_t$ is converted from RGB color space to the brightness channel Y (from YCbCr color space) and replicated 16 times to match the residual's dimension. This procedure is effectively nearest-neighbor interpolation. No information is altered during this process, which means the network in parallel does not need to allocate complexity to learn this transformation, as it would be the case with other methods, \eg bicubic interpolation. In contrast to FRVSR, which does not adopt this strategy, all complexity can be used to reconstruct only the missing high-frequency components. All LR inputs and nearest-neighbor interpolated HR output (represented in LR space) are properly aligned. 

\subsection{Feedback}
Feeding back the output naturally helps to improve continuity between frames and reduces flickering, which can occur in models with limited temporal connectivity. Because of high correlation between adjacent frames, having a reference of the previous output $y_{t-1}$ also supplies additional, already processed HR information when producing the HR estimate $y_t$. 

\subsection{Hidden State}
In order to propagate complex, abstract information across time, a hidden state $h_t$ is added to the processing chain. The hidden state is realized by carrying forward the feature maps from the previous iteration and feeding them back to the input through concatenation. This keeps the network structure in line with the prior on locality.
Since the whole network is fully convolutional, the hidden state's spatial dimensions are dynamically adjusted according to the input size of $x$. The hidden state can be seen as a set of vectors in a locality based latent space $\mathbb{R}^f$, characterized by vectors $v \in \mathbb{R}^{f}$ with entries along the channel axis, see Fig.~\ref{fig:latentspace}. Since every instance in a feature map is calculated by the same convolution kernel, each feature map represents a dimension in the latent space $\mathbb{R}^{f}$. In contrast to using just feedback as done in~\cite{frvsr}, which is bound to pass HR information from $y_{t-1}$ only, the hidden state is not limited in the type of information that can be propagated. It is theoretically possible to propagate past information across the whole time axis.

Because at time instance $t$, the next frame $x_{t+1}$ is fed to the input already, the hidden state also allows every frame $x_t$ to be processed twice before estimating $y_t$. This essentially increases the receptive field and the number of processing layers for a single frame, even though both processing steps share their weights. Because of recurrence, these two steps can be efficiently distributed over two time steps. 

\begin{figure}
\begin{center}
\includegraphics[width=0.50\linewidth]{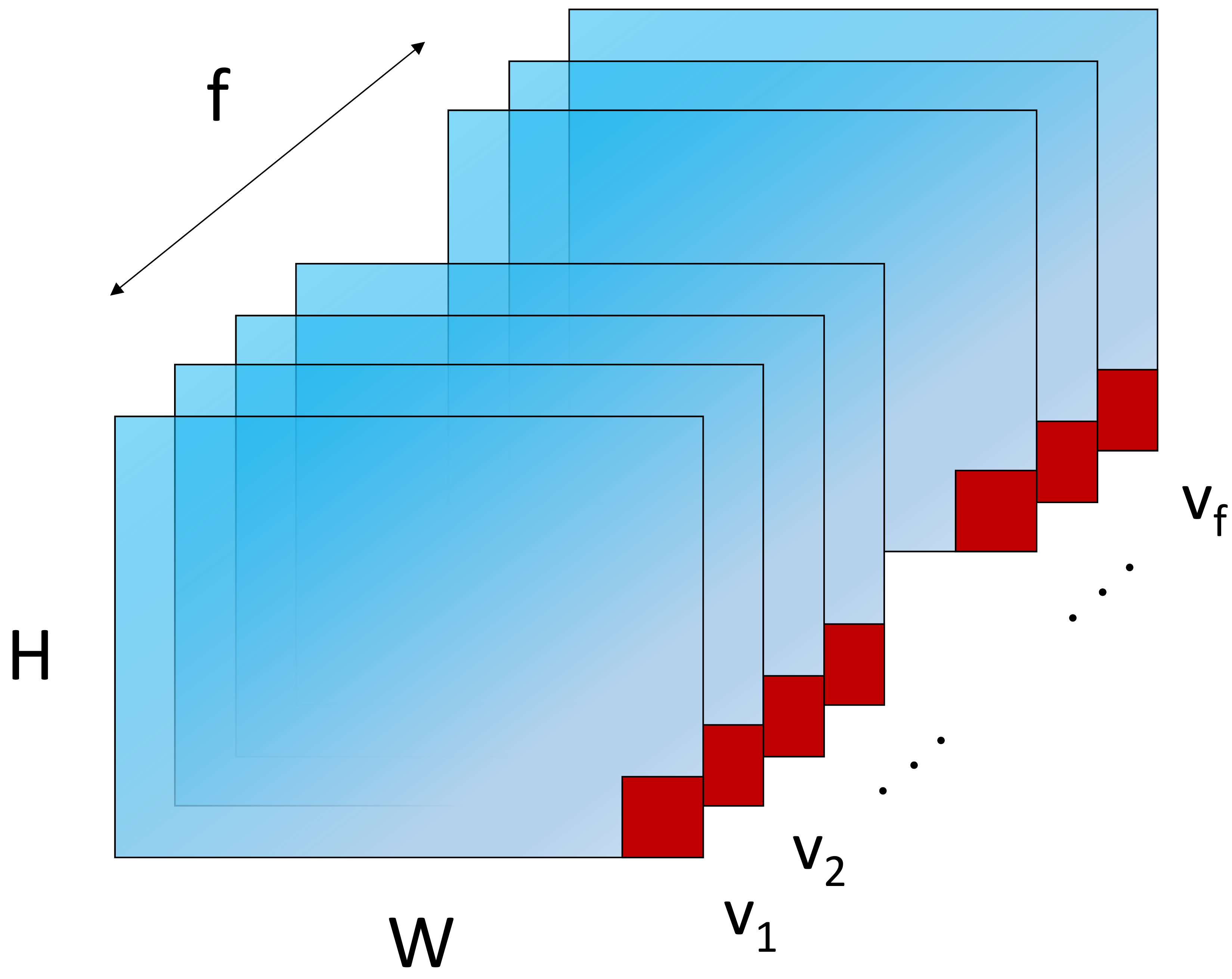}
\end{center}
\vspace{-0.14cm}
   \caption{Locality based hidden state. Entries along the channel dimension represent vectors $v \in \mathbb{R}^f$.}
\label{fig:latentspace}
\vspace{-0.14cm}

\end{figure}

\subsection{Loss}
The loss function is defined as the pixel-wise mean-squared-error (MSE) between all $k$ pixels in the ground truth frames $y^{*}$ and the network's output $y$:
\begin{equation}
\mathcal{L} = \frac{1}{k}\norm{y^{*} - y}^2_2
\end{equation}
\section{Experimental Setup}

\subsection{Dataset}
We adhere to the experimental setup from ~\cite{frvsr} and use the same dataset. Originally, it contained 40 high resolution videos (720p, 1080p, 4k), downloaded from \textit{vimeo.com}, but 3 videos were not online anymore, so we train on the available 37 videos instead. Following the same procedure as in~\cite{frvsr} we produce 40,000 random crops of size $20\times 256\times 256 \times 3$, which serve as HR ground truth sequences $y^* \in \mathbb{R}^{T\times H \times W \times C}$. To get the corresponding LR sequences $x$, Gaussian blur with $\sigma = 1.5$ is applied and every 4-th pixel in both spatial dimensions is sampled as done in both methods~\cite{frvsr, duf} that we use for comparison.
\newline
To monitor training progress and generalization, we downloaded 10 additional high resolution videos from \textit{youtube.com} and generated validation sequences with the same procedure as used for generating the training data.

\subsection{Training}
\label{sec:training}
To train our RNN, the cell is unrolled along the time axis to accommodate for a training clip length of 10 frames. For training, we randomly sample 12 consecutive frames from the training clips, which contain 20 frames. The 2 additional frames are used to feed $x_{t-1}$ at the beginning and $x_{t+1}$ at the end. The weights are initialized with Xavier initialization~\cite{xavier} and the network is trained with batches of size 4 with Adam optimizer~\cite{adam}. Because of the network's recurrent structure, the hidden state $h_{t-1}$, and the previous estimate $y_{t-1}$ need to be initialized. Both tensors are initialized with zeros. In our experimental setup, our fastest model has 7 layers with 48 filters per layer, denoted as RLSP 7-48, while our most accurate model is implemented with 7 layers and 256 filters (RLSP 7-256). 
The networks are trained with decreasing learning rate, starting at $10^{-4}$. For RLSP 7-48 the learning rate is divided by 10 after 2M and 3M iterations. For all other models, the learning rate is divided by 10 after 2M and 4M steps. The models are selected according to the lowest moving average on the validation loss at convergence.

\section{Results and Discussion}

We investigate our models in terms of runtime, accuracy, temporal consistency, information flow and provide images for qualitative comparison. We compare our models with state-of-the-art video super-resolution methods DUF~\cite{duf} and FRVSR~\cite{frvsr} on Vid4~\cite{bayes} benchmark. To the best of our knowledge DUF achieves the highest accuracy while FRVSR is very efficient with the best tradeoff between accuracy and runtime to date.

\subsection{Ablation}

We compare different configurations found in other VSR methods by applying them to our architecture and assess the impact of each part in terms of accuracy and runtimes. A SISR implementation of our network serves as a baseline. For that matter, all recurrent connections are removed and only the current frame $x_t$ is fed to produce $y_t$. As an extension to SISR, a batch of 3 consecutive frames $x_{t-1:t+1}$ is fed to the network. This approach is further expanded by introducing a recurrent feedback connection $y_{t-1}$. Finally, our proposed locality based hidden state is added. All configurations are trained on the same data with the settings, described in Sec.~\ref{sec:training}. The experiments are conducted with $n=7$ layers and $f=64$ filters. The PSNR values on Vid4 and Full HD runtimes are shown in Tab.~\ref{tab:ablation}.

Adding adjacent frames already improves PSNR substantially compared to the SISR implementation by 1.3dB. Feeding back the previous output improves PSNR by 0.23dB. Our proposed locality based hidden state further improves the PSNR score substantially by 0.55dB. The experiments show, that adjacent frames and our locality based hidden state, have the strongest impact on the video performance.
As expected, the runtime increases for more complex configurations. The models without recurrence need 12ms to produce a single Full HD frame. Adding feedback ($y_{t-1}$) and the hidden state ($h_{t-1}$) to the network increases runtime by 2ms and another 5ms, respectively. The complete configuration RLSP 7-64 achieves a PSNR value comparable to recent state-of-the-art by gaining 1.98dB compared to the SISR implementation. RLSP 7-64 can generate 50fps Full HD in real time.

\begin{table}[h!]
\begin{center}
\begin{tabular}{|l|c|c|}
\hline
Inputs & PSNR [dB] & Runtime [ms] \\
\hline\hline
$x_t$ & 24.91 & 12 \\
$x_{t-1:t+1}$ & 26.20 & 12 \\
$x_{t-1:t+1}$ + $y_{t-1}$ & 26.43 & 14 \\
$x_{t-1:t+1}$ + $y_{t-1}$ + $h_{t-1}$ & 26.89 & 19 \\
\hline
\end{tabular}
\end{center}
\vspace{-0.14cm}
\caption{Results for different network configurations on Vid4.}
\label{tab:ablation}
\vspace{-0.14cm}
\end{table}

\subsection{Temporal Consistency}

\begin{figure*}
\begin{center}
\includegraphics[width=1\linewidth]{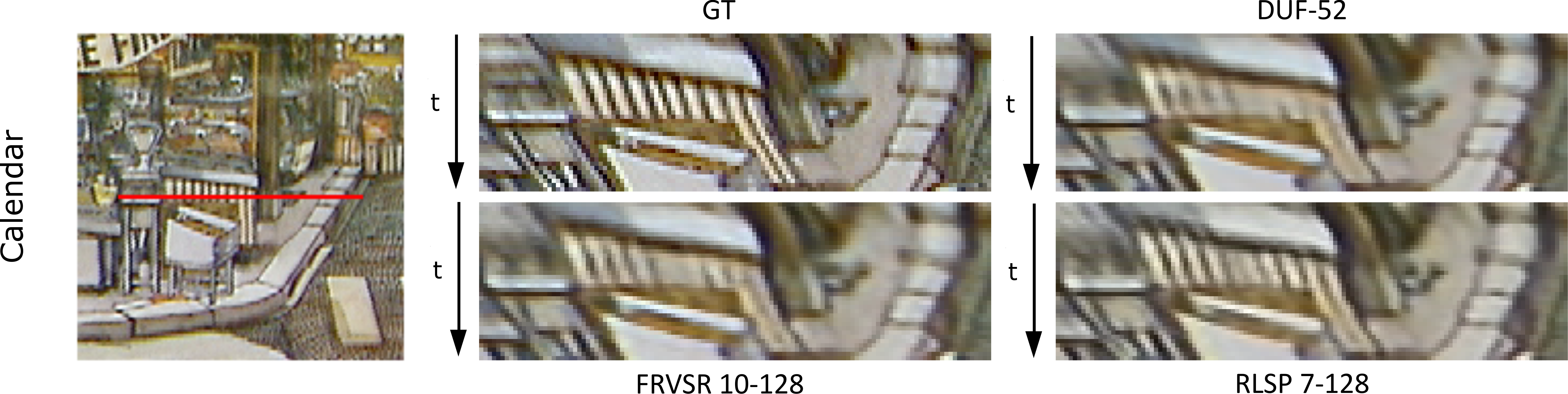}
\end{center}
   \caption{Temporal profiles for calendar. The profiles are produced from the red line, shown on the left}
\label{fig:temp_profile}
\end{figure*}

An important aspect in VSR is temporal consistency between consecutive frames. Methods with limited temporal connectivity often expose flickering and other artefacts along the time domain. This property can not be analyzed by per frame PSNR values. Therefore, we provide temporal profiles to visually assess the temporal continuity of our RLSP method compared to others. For that matter, a single pixel line (red line in Fig.~\ref{fig:temp_profile}) is recorded along the whole sequence and stacked vertically. Temporal profiles of high quality videos with smooth temporal transitions expose sharp detailed images. 
DUF-52 is the most limited method in terms of exploiting temporal information, as it only uses adjacent frames without any recurrent connections. FRVSR 10-128 uses feedback to leverage temporal information. Our method additionally propagates a latent state, which increases temporal connectivity even further.
These properties are also reflected in the temporal profiles in Fig.~\ref{fig:temp_profile}. The vertical stripes in DUF-52's profile are blurred out, which indicates discontinuities between consecutive frames. FRVSR 10-128 shows increased performance, while our method RLSP 7-128 exhibits the sharpest stripes among all methods. Our method can therefore produce better temporal consistency overall.

\subsection{Information Flow over Time}

\begin{figure*}[th!]
\begin{center}

\includegraphics[width=\linewidth]{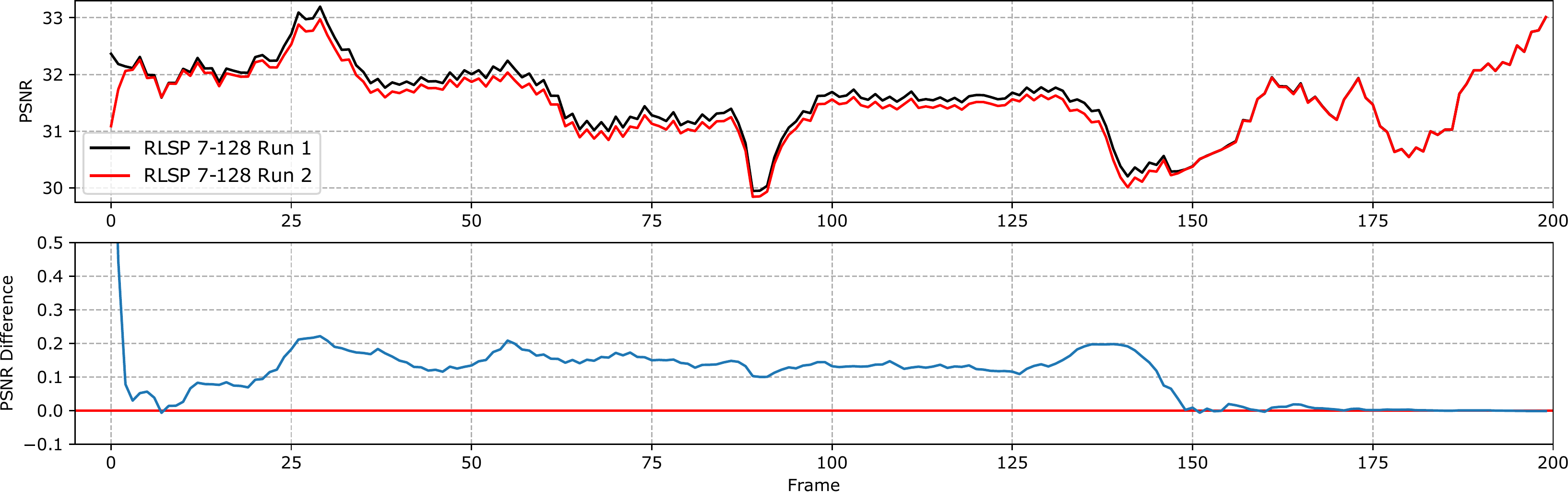}
\end{center}
   \caption{Information flow over time for model RLSP 7-128 on a validation video. Top: Absolute PSNR values, Bottom: Difference in PSNR per frame.}
\label{fig:catchup}
\end{figure*}

Unlike many existing methods, our model is not limited by the number of LR input frames to extract information over time. To investigate the range of information flow, the same model (RLSP 7-128) is evaluated on a Full HD sequence from the validation set, but initialized at different instances in time. The first run is started 100 frames ahead of the second run. Therefore, the first run has already accumulated information over 100 frames, when the second run is initialized at frame 0. The respective PSNR values per frame and the difference between the two runs are plotted for 200 frames in Fig.~\ref{fig:catchup}.
The experiment shows that the model can propagate information over almost 175 frames until the two runs finally converge. Accumulated information from the first 100 frames can be leveraged to get up to 0.2dB higher PSNR values over a long period of 150 frames. Interestingly, the two runs collapse at the beginning, but quickly separate again, which leads to the conclusion, that the model saves information, based on considering a large horizon.

\subsection{Initialization}

\begin{figure*}
\begin{center}
\includegraphics[width=1\linewidth]{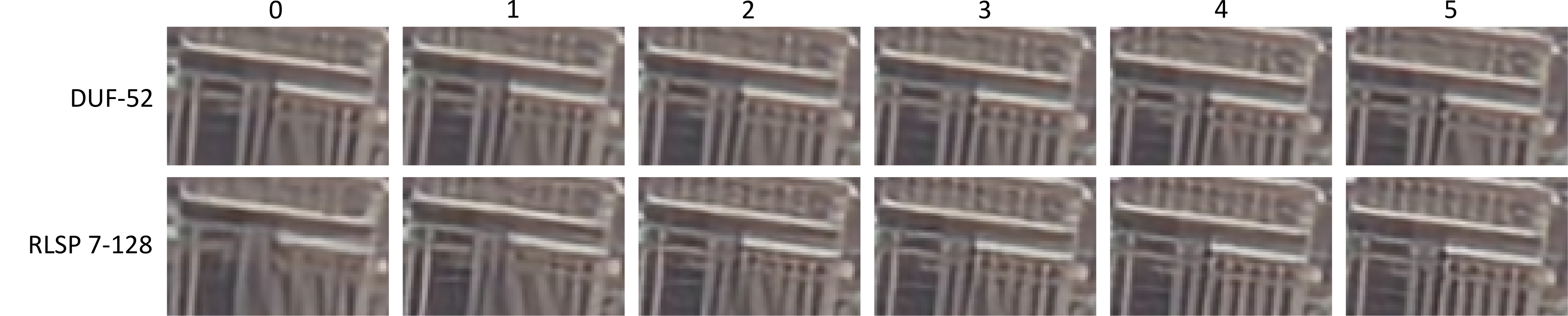}
\end{center}
   \caption{Initialization artefacts for DUF-52 and RLSP 7-128. The first 6 frames of \textit{city} are shown. DUF-52 is fully initialized at frame 2, when all input frames $x_{t-2:t+2}$ are available. RLSP 7-128 reconstructs the full structure starting from frame 4, while DUF-52 still exhibits artefacts.}
\label{fig:rampup}
\end{figure*}

Due to the recurrent nature, our method is dependent on previously processed information. Because the available information content is at its lowest at the beginning, it takes a couple of frames to gather temporal information. This phenomenon can be observed in Fig.~\ref{fig:rampup}, where the first 6 frames of sequence \textit{city} are shown. We compare our method RLSP 7-128 with DUF-52, which suffers from initialization until frame 2 as well.
The fine structure of the building in the first frame can not be fully reconstructed, but as more information is processed, the quality increases quickly until the correct structure is revealed after frame 4. This behaviour is also represented in the PSNR values in Fig.~\ref{fig:vid4_average}. Our models start with lower PSNR at the beginning, but quickly catch up with DUF to then surpass it. Unfortunately, the first 2 and the last two frames of FRVSR 10-128 are not provided by the authors, probably, because these frames are not considered for PSNR evaluation.

\subsection{Accuracy and Runtimes}

\begin{figure*}
\begin{center}
\includegraphics[width=\linewidth]{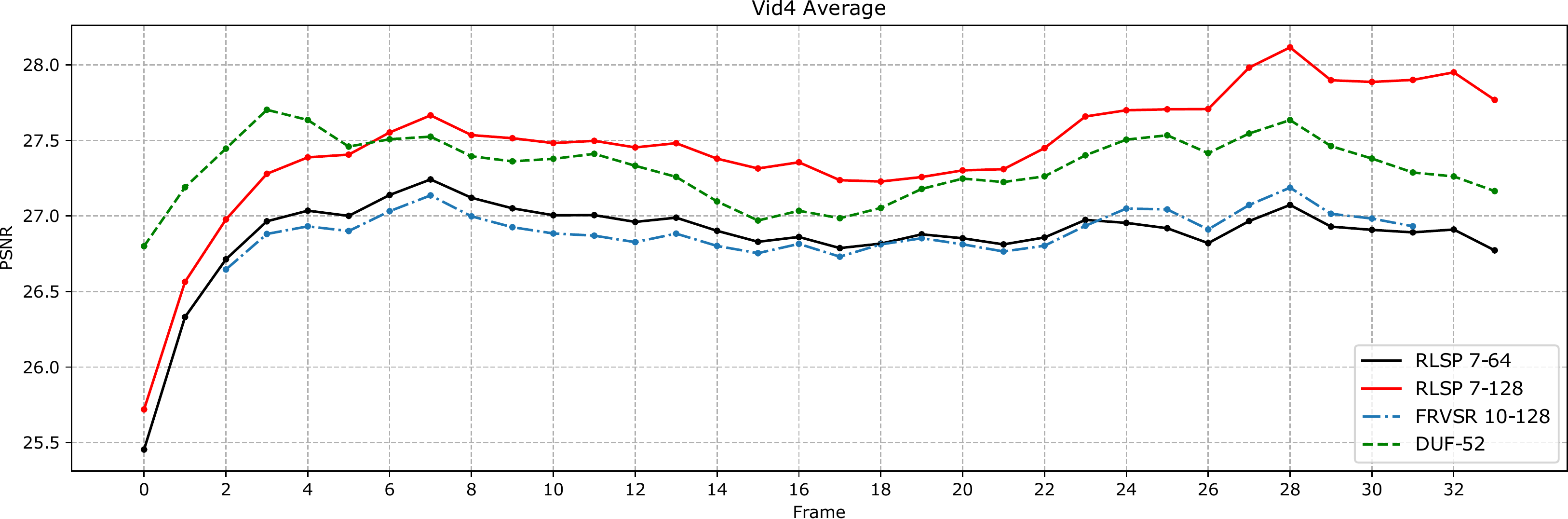}
\end{center}
\vspace{-0.15cm}
   \caption{Average PSNR values over all 4 videos in Vid4 for the top models of FRVSR~\cite{frvsr}, DUF~\cite{duf} and our models RLSP 7-64 and RLSP 7-128. The full average is calculated for the first 34 frames, constrained by the shortest sequence \textit{city}. The first and last two frames of FRVSR 10-128 are not provided by the authors.}
\label{fig:vid4_average}
\vspace{-0.15cm}
\end{figure*}

To compare performance of the proposed network with other methods we calculate the average PSNR over all sequences from Vid4 and measure runtimes to produce a single Full HD (1920x1080) frame. The ideal method attains fast runtimes and high PSNR.
Video PSNR is calculated from the MSE over the total number of pixels in a single video sequence. The final reported value is the average of each sequence's PSNR value. Sometimes, methods use encoder-decoder parts in their architectures and therefore need to restrict input sizes. It is therefore common to crop the spatial dimensions on test sets to accommodate for this drawback. Our method can process any input dimension and cropping would not be required. However, to objectively compare our method to others, we follow the evaluation strategy described in~\cite{duf} and directly include the reported values from that paper. The values for FRVSR 10-128~\cite{frvsr} are recalculated in the same way from the provided output images. Because the outputs for model FRVSR 3-64 are not provided, we simply add the same difference in PSNR on top of the reported value in the paper, which was gained, when calculating the new PSNR value for FRVSR 10-128. Our runtimes are measured on a NVIDIA TITAN Xp with our unoptimized code. DUF and FRVSR runtimes are taken from the respective papers. The results are listed in Tab.~\ref{tab:results} and displayed in Fig.~\ref{fig:t_vs_acc}.

\begin{figure*}[th!]
\begin{center}
\text{\hphantom{aaaiiaaa}GT 
\hphantom{aaaaaaaaaiaaa} Bicubic 
\hphantom{aiiaaa}  FRVSR 10-128 \cite{frvsr} 
\hphantom{aaa} DUF-52 \cite{duf} 
\hphantom{aaaaiaa} RLSP 7-64 
\hphantom{aaaiaaa} RLSP 7-128 
\hphantom{aaaaaaaaaaaaaaaaaaaaaaaaaaaaaaaaaaaaaaaaaaaaaaaaaaaaaaaaaaaaaaaaaaaaaaaaaaaaaaaaaaaaaaaaaaa}} \par\medskip
\includegraphics[width=\linewidth]{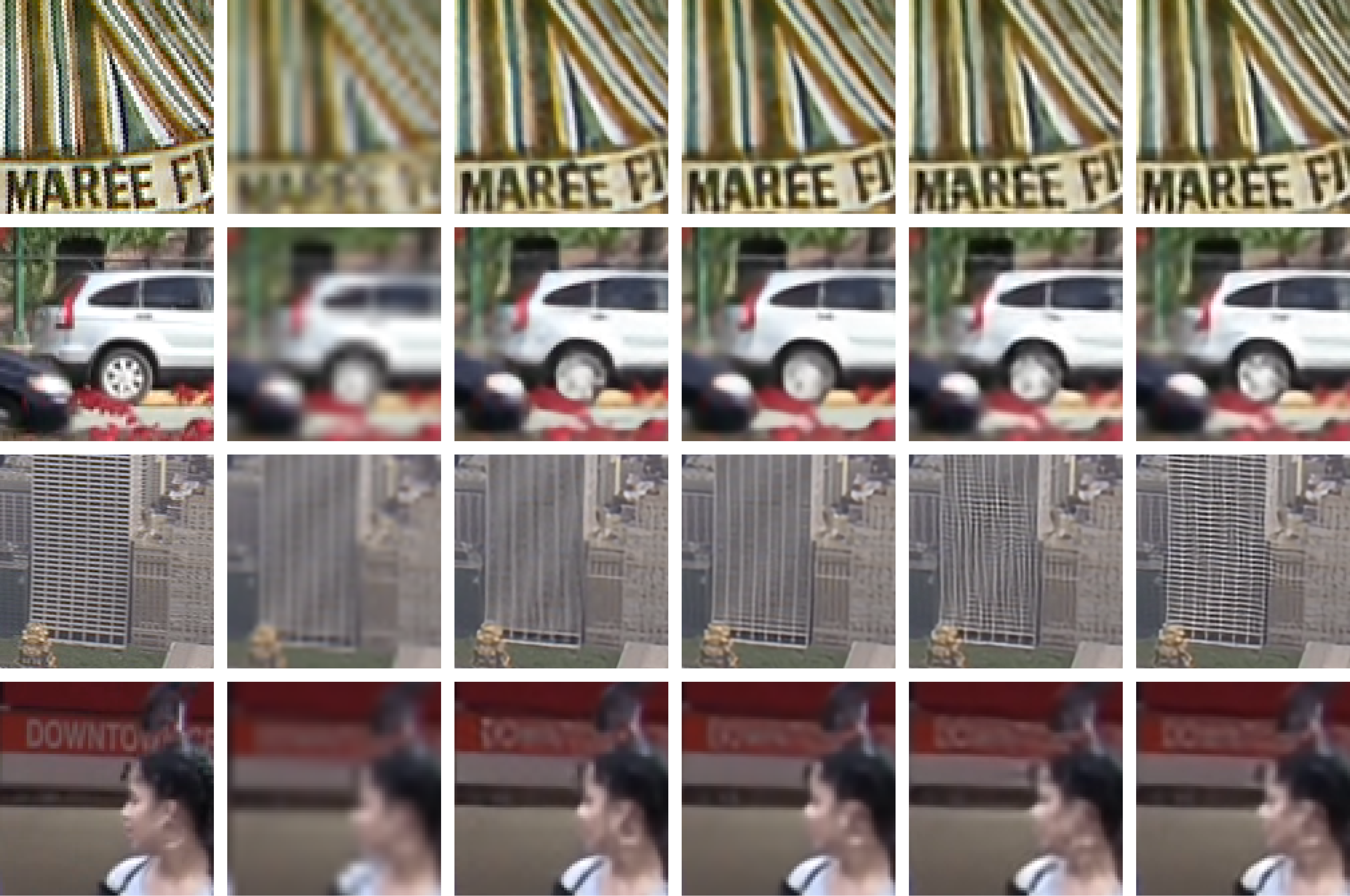}
\end{center}
\vspace{-0.15cm}
   \caption{Visual comparison on Vid4. From top to bottom: \textit{calendar}, \textit{foliage}, \textit{city}, \textit{walk}.}
\label{fig:vid4_comparison}
\end{figure*}

\begin{table*}[th!]
\begin{center}
\begin{tabular}{|c||c||c|c||c|c|c||c|c|c|c|}
\hline
Method & Bicubic &  \begin{tabular}{@{}c@{}} FRVSR \\ 3-64 \end{tabular} & \begin{tabular}{@{}c@{}} FRVSR \\ 10-128 \end{tabular} & \begin{tabular}{@{}c@{}} DUF \\ 16 \end{tabular} & \begin{tabular}{@{}c@{}} DUF \\ 28 \end{tabular} & \begin{tabular}{@{}c@{}} DUF \\ 52 \end{tabular} & \begin{tabular}{@{}c@{}} RLSP \\ 7-48 \end{tabular} & \begin{tabular}{@{}c@{}} RLSP \\ 7-64 \end{tabular} & \begin{tabular}{@{}c@{}} RLSP \\ 7-128 \end{tabular} & \begin{tabular}{@{}c@{}} RLSP \\ 7-256 \end{tabular} \\
\hline\hline
PSNR [dB] & 23.79 & 26.38 & 26.90 & 26.81 & 26.99 & 27.34 & 26.65 & 26.89 & 27.46 & \textbf{27.55} \\
\hline
Runtime [ms] & - & 74 & 191 & 403 & 838 & 2819 & 15 & 19 & 38 & 92\\

\hline
\end{tabular}
\end{center}
\caption{Comparison of PSNR values on Vid4 and runtimes between methods FRVSR \cite{frvsr}, DUF \cite{duf} and ours. Bicubic interpolation is included as a baseline. All runtimes are computed on Full HD.}
\label{tab:results}
\vspace{-0.2cm}
\end{table*}

RLSP 7-64 shows comparable PSNR to FRVSR 10-128 and DUF-16, but is 10$\times$ and 20$\times$ faster, respectively. RLSP 7-128 achieves a good trade-off between accuracy and speed. It reaches state-of-the-art accuracy, while reducing runtimes by two orders of magnitude compared to DUF-52. Our direct implementation without any optimization can produce 25fps of Full HD video in real-time. 
Because the focus in this work is on performance, we keep the layer count constant at 7 and instead vary the number of filters. Due to the high parallelizability, this allows to increase complexity without putting too much burden on computation time. By further increasing the number of filters, RLSP 7-256 achieves the highest reported PSNR to date on Vid4, improving 0.65dB over FRVSR 10-128 and 0.21dB over DUF 52 while still being more than $2\times$ faster than FRVSR and $30\times$ faster than DUF, respectively.

To investigate evolution of accuracy across time, average PSNR per frame over all sequences in vid4 is computed and plotted in Fig.~\ref{fig:vid4_average}. All methods have to deal with incomplete initialization. Since our RLSP approach profits greatly from past information, PSNR values are lower at the beginning compared to DUF-52. However, RLSP 7-128 improves quickly and is able to surpass all other methods from frame 6 until the end.

We also provide images for visual comparison in Fig.~\ref{fig:vid4_comparison} for each sequence in Vid4.

\section{Conclusion}

We introduced RLSP, a new end-to-end trainable recurrent video super-resolution architecture with locality based latent space propagation, without relying on a dedicated motion estimation module. Due to the ability of effectively leveraging temporal information over long periods of time, RLSP reduces runtimes drastically, while still maintaining state-of-the-art accuracy.
Because our network is shallow and wide, a large amount of computation can be run in parallel, which is also responsible for its efficiency and could enable even faster runtimes for dedicated hardware implementations. 
Even though the network structure is designed to be highly efficient, we could show, that it is still possible to improve accuracy by increasing complexity, \eg adding more filters. Our RLSP achieves the best accuracy on Vid4 benchmark while being more than $70\times$ faster than DUF, the former state-of-the-art. Accuracy could also be further improved by investigating different configurations of kernel sizes, alternative convolution types or numbers of layers.


{\small
\bibliographystyle{ieee}
\bibliography{egbib}
}

\end{document}